\documentclass[final,3p,times,twocolumn]{elsarticle}

\usepackage{amssymb}
\usepackage{amsmath,cleveref}
\usepackage{float}
\usepackage{centernot}
\usepackage{comment}
\usepackage{caption}
\usepackage{subcaption}

\usepackage{algorithm}
\usepackage{algpseudocode}

\usepackage{tikz}
\usetikzlibrary{shapes, arrows}
\usetikzlibrary{positioning, arrows.meta, fit, shapes}

\usepackage{tabularx}
\usepackage{colortbl}  
\usepackage{geometry}  

\usepackage{graphicx}
\usepackage{verbatim}

\usepackage{tikz}
\usetikzlibrary{positioning, arrows.meta}
\usetikzlibrary{positioning, decorations.pathreplacing}

\newcounter{definition}

\newcommand{\definition}[2]
{\refstepcounter{definition}%
\par\noindent \newline \textbf{Definition}  \thedefinition\ (\emph{#1}):  #2\medskip\par}

\begin{document}

\begin{frontmatter}



\title{
	A Fusion-Driven Approach of Attention-Based CNN-BiLSTM for Protein Family Classification - ProFamNet
}






\author[imsciences]{Bahar Ali}
\affiliation[imsciences]{organization={School of Computer Science and IT, Institute of Management Sciences},
	city={Peshawar},
	country={Pakistan}}

\author[fast_nuces2]{Anwar Shah \corref{cor1}}
\affiliation[fast_nuces2]{organization={National University of Computer and Emerging Sciences},
	city={Faisalabad},
	country={Pakistan}}

\author[fast_nuces]{Malik Niaz}
\affiliation[fast_nuces]{organization={National University of Computer and Emerging Sciences},
	city={Peshawar},
	country={Pakistan}}

\author[gik]{Musadaq Mansoor}
\affiliation[gik]{organization={Ghulam Ishaq Institute of Emerging Sciences},
	city={Topi},
	country={Pakistan}}

\author[fast_nuces]{Sami Ullah}

\author[fast_nuces2]{Muhammad Adnan}
\cortext[cor1]{Corresponding author. Email: anwar.shah@nu.edu.pk}

\begin{abstract}
Advanced automated AI techniques allow us to classify protein sequences and discern their biological families and functions. Conventional approaches for classifying these protein families often focus on extracting N-Gram features from the sequences while overlooking crucial motif information and the interplay between motifs and neighboring amino acids. Recently, convolutional neural networks have been applied to amino acid and motif data, even with a limited dataset of well-characterized proteins, resulting in improved performance. This study presents a model for classifying protein families using the fusion of 1D-CNN, BiLSTM, and an attention mechanism, which combines spatial feature extraction, long-term dependencies, and context-aware representations. The proposed model (ProFamNet) achieved superior model efficiency with 450,953 parameters and a compact size of 1.72 MB, outperforming the state-of-the-art model with 4,578,911 parameters and a size of 17.47 MB. Further, we achieved a higher F1 score (98.30\% vs. 97.67\%) with more instances (271,160 vs. 55,077) in fewer training epochs (25 vs. 30).

%

\end{abstract}



\begin{keyword}
Protein Family Classification · Deep learning · Convolutional neural network · Long short-term memory · Attention network




\end{keyword}

\end{frontmatter}


\pagebreak 

\section{Introduction}
Examining relationships among smaller units in a protein sequence, like amino acids and motifs, provides deeper insights into the functional characteristics of associated physical entities. Identifying distinguishing amino acid residues among protein families is a dynamic research focus in biomolecular recognition. The researchers classify protein families as clusters of proteins sharing analogous functions. However, a significant proportion of proteins remains uncharacterized compared to their well-studied counterparts in the field of bioinformatics research. Consequently, the challenge of classifying proteins and clarifying their functionalities solely based on their physiochemical properties is a crucial concern in this field, particularly when it comes to annotating proteins that have not been previously studied\citep{nguyen2016hippi}. 

In the traditional approach to protein sequence analysis, the process involves employing feature engineering techniques to manually extract either discrete or continuous features. Conventional machine learning algorithms are then deployed to extract underlying patterns within these features. One common method among these conventional machine learning approaches is unsupervised clustering, which is widely utilized to create clusters and assign labels to each of these clusters\citep{vijaya2006efficient, boujenfa2011tree, dongardive2016protein, chappell2017k}. Another prevalent technique involves scrutinizing protein sequences by searching for common motifs or aligning genetic characteristics with known protein sequences\citep{elayaraja2012extraction, livingstone1993protein}. However, this process of comparing common motifs relies heavily on the expertise of biological professionals and domain-specific knowledge to identify functional motifs.

Recent studies employ Neural Networks (NN) computational models, achieving state-of-the-art performance in addressing complex problems. This remarkable proficiency has opened up the opportunity to integrate NN-based architectures also into various domains within the field of biology. Some of these applications draw inspiration from the Natural Language Processing (NLP) field\citep{goldberg2022neural, karuppusamy2020analysis, widiastuti2019convolution}, while others are derived from deep neural network architectures\citep{torfi2020natural, dai2021deep, bharadiya2023comprehensive}.

In neural network-based architectures, 1D Convolutional Neural Networks (1D-CNNs) have recently entered Natural Language Processing (NLP), excelling in tasks such as text classification and machine translation\citep{soni2023textconvonet, nouhaila2021arabic}. Besides their conventional role in image classification and object captioning, CNNs have even been extended to process sequence data, including DNA/RNA binding sequences. In contrast to RNN networks, 1D-CNNs have the potential to enhance their performance by incorporating multiple convolutional layers, thereby expanding their receptive fields. 1D-CNN with fully connected layers have also been used for the protein functions classification producing outstanding results\citep{alipanahi2015predicting, gunasekaran2021analysis, patiyal2023prediction}. LSTM-based models are designed with the inclusion of extra  ``$gated$" to facilitate the retention of longer input data sequences. These gates within the LSTM architecture significantly enhance predictive capabilities. In contrast, Bidirectional LSTMs (BiLSTMs) further enhance model training by processing the input data in both forward and backward directions. The outcomes of this approach demonstrate that the additional training provided by BiLSTMs yields superior predictive performance when compared to conventional LSTM-based models\citep{siami2019performance}. CNN and LSTM with attention have also been applied in NLP e.g for subjectivity detection in opinion mining\citep{sagnika2021attention}. The authors proposed a fusion-based approach, combining MRI and demographic features, which significantly enhances the performance of Alzheimer's disease diagnosis. This recent trend demonstrates a promising improvement in diagnostic accuracy \citep{rahim2023prediction}. 

This research is motivated by the fusion of 1D-CNNs and LSTM with attention, highlighting their success in NLP, along with the performance advantage of BiLSTM over LSTM. The authors introduce a comprehensive deep learning framework using the fusion of 1D-CNN, BiLSTM, and an attention mechanism for the classification of protein families. The authors refer to this approach as the ``Attention-Based CNN-BiLSTM for Protein Family Classification", which is ``ProFamNet" in short. This process entails initially encoding each amino acid as a numerical value and subsequently representing it as a vector with a predefined length. Once each sequence is embedded as a matrix, we input this data into the deep 1D CNN with dropout mechanism and maxpooling followed by BiLSTM with attention network. The network then extracts profound and abstract representations from the input sequence data resulting in the  effective performance of protein classification tasks. The key contributions of the paper are as follows;

\begin{itemize}
	\item Introduced a novel model namely ProFamNet for protein family classification.
	\item Utilized 1D-CNN combined with BiLSTM and attention mechanisms.
	\item Achieved superior model efficiency with fewer parameters and more compact size. 
	\item Surpassed the baseline in terms of fewer layers and training epochs, significantly lowering the training time and resource consumption.	
	\item Outperformed state-of-the-art models in terms of F1 score across a broad spectrum of labels, highlights the effectiveness of the model in bioinformatics.
\end{itemize}

\section{Related works and paradigms}
\begin{figure*}[t]
	\centering
	\includegraphics[width=0.80\textwidth]{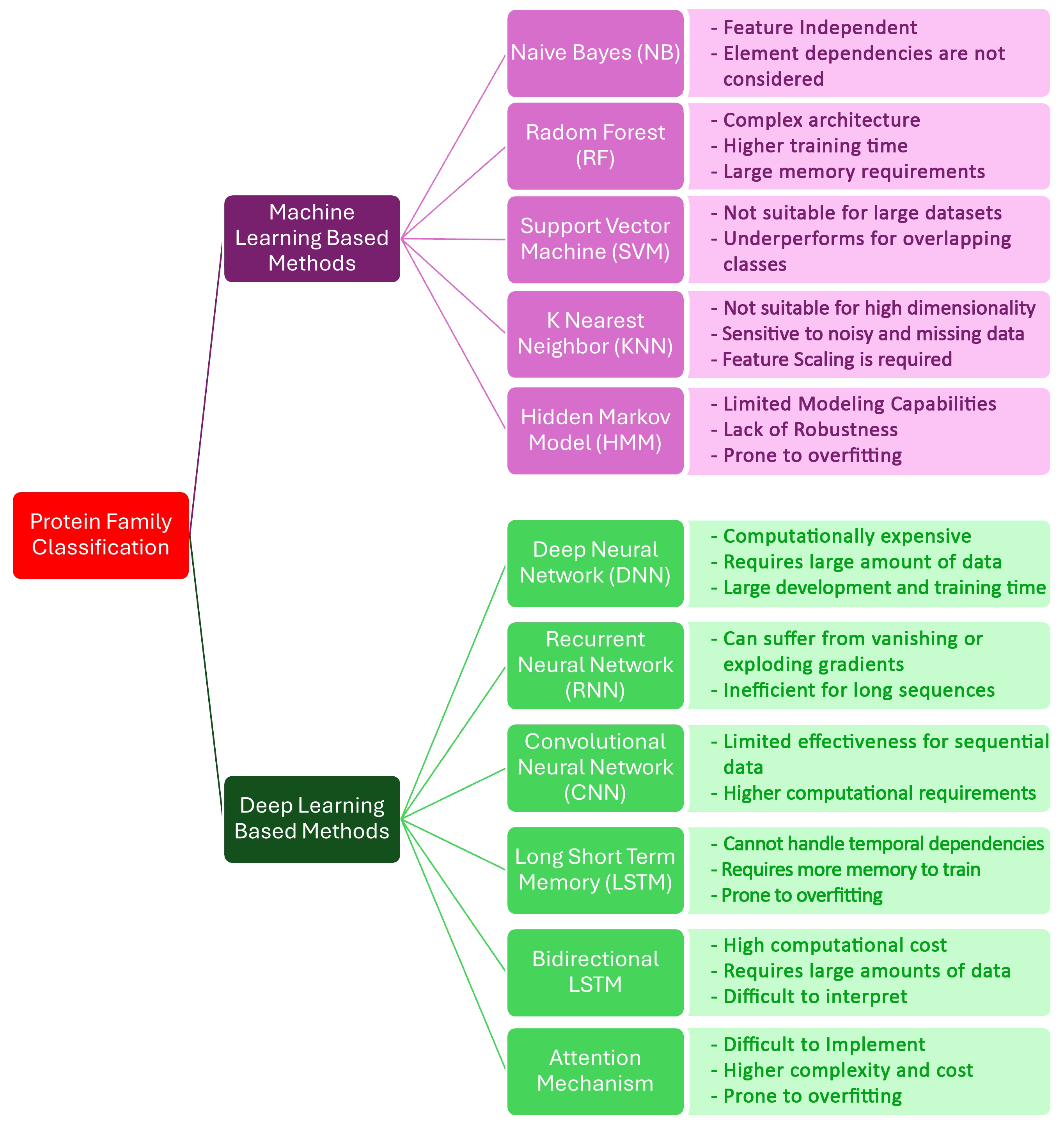}
	\caption{Literature review }
	\label{fig:Literature-Review-Final}
\end{figure*}

In this section, we review the techniques used for classifying biological sequences within the framework of both machine learning and deep learning. We cover both traditional machine learning techniques and the latest advancements in deep learning. A key initial step is feature extraction, where relevant attributes are crafted for machine learning. In the context of biological sequences, like DNA or protein data, feature engineering typically involves human expert knowledge. To address this reliance, automated feature learning techniques are also reviewed for deriving meaningful representations directly from data. This is particularly helpful in handling the complexities of protein sequences. The review also  provides an overview of the current state of the field, including the latest automated tools for analyzing biological sequences. For a quick review, refer to Figure \ref{fig:Literature-Review-Final}, which pictorially illustrates the research conducted in this field and highlights the gaps that need to be addressed.

\subsection{Traditional machine learning techniques}\label{Traditional_machine_learning}
These techniques involve features extraction for learning and producing classes from the amino acid sequences. Some examples of traditional machine learning are Naive Bayes (NB), Random Forest (RF), Support Vector Machine (SVM), K-Nearest Neighbor (KNN) and Hidden Markov Model (HMM). A detail of each of these is given in the following as a ready reference for the reader.

\subsubsection{The Naive Bayes (NB)}
The Naive Bayes machine learning technique is a probabilistic classification method based on Bayes' theorem. It is particularly known for its simplicity and efficiency in handling classification tasks. Naive Bayes assumes that features are independent, and conditional probabilities among these are learned in each training step. Despite this simplification, it often performs surprisingly well in various applications, ranging from text classification tasks to gnome sequences analysis\citep{cheng2005protein}. However, it faces a prevalent constraint in practical use due to its assumption of feature independence. In real-world situations, elements in sequences frequently display dependencies that are not considered by the Naive Bayes model.

\subsubsection{The Radom Forest (RF)}
The Random Forest is a powerful machine learning technique. It operates by constructing several decision trees during the training phase with each tree created on a random subset of the dataset. This inherent randomness and diversity in the construction of individual trees avoid over-fitting and enhance generalization. In the prediction phase, the model aggregates the results from these multiple trees to make robust and accurate predictions. Random Forests is used in a wide range of applications, including classification, regression, and feature selection. Its ensemble nature leads to some methods to extract multi-scale local descriptor (MLD) as various lengths of amino acid segments for predicting protein-protein interactions\citep{you2015predicting}.

\subsubsection{The Support Vector Machine (SVM)}
An effective approach for the analysis of protein sequences involves utilizing a Support Vector Machine (SVM) based method was introduced by Shen et al \citep{shen2007predicting}. In their study, 20 amino acids are systematically classified into seven distinct classes, considering their unique dipoles and side chain volumes. Subsequently, this method derives conjoint triad features from protein pairs, utilizing various amino acid types. Shen et al. primarily applied their research to forecast human Protein-Protein Interactions (PPIs) and achieved a prediction accuracy of 83.9$\%$.

Building upon the work of Shen et al, Guo et al. presented an approach rooted in SVM and auto-covariance to extract vital interaction information from discontinuous segments of amino acids within protein sequences \citep{guo2008using}. By employing this technique, the accuracy of PPI predictions for the S.cerevisiae species rose to 86.55$\%$.

\subsubsection{The K Nearest Neighbor (KNN)}
The K Nearest Neighbor (KNN) approach is a technique rooted in sequence similarity analysis, particularly focusing on defining various distance functions such as Euclidean distance, dynamic time warping distance, and alignment distance. These functions serve to quantify the similarity between a pair of sequences. In practical terms, when presented with a labeled sequence dataset and a predetermined number of nearest neighbors, denoted as `$k$' and a new sequence query `$s$' awaiting classification, the KNN method tries to identify the `$k$' most similar neighbors to `$s$' within the training dataset `$T$'. It then employs a voting algorithm to determine the dominant class labels among these neighbors\citep{kajan2006application}.

Among the available distance functions, alignment-based distance measurement is popular in the context of labeling unlabeled proteins. By adopting this approach, the collective knowledge contained within labeled sequences is utilized to make informed decisions about newly arriving sequences.

\subsubsection{The Hidden Markov Model (HMM)}
The Naive Bayes model, known for its simplicity in implementation, encounters a common limitation in practical applications due to its assumption of feature independence. In real-world scenarios, elements within sequences often exhibit dependencies that are not accounted for by NB. To tackle this issue, the Hidden Markov Model (HMM) provides a valuable framework to capture and consider these dependencies. A notable advancement in this direction is the work of Yakhnenko and colleagues, who have applied the concept of the k-order Markov model. This model allows for the classification of protein sequences while taking into account the inherent dependencies between sequence elements\citep{yakhnenko2005discriminatively}.

\subsection{Deep Learning Methods}
Deep learning is a subset of machine learning that uses artificial neural networks with multiple layers (deep neural networks) to automatically learn and extract complex patterns from data, making it applicable for tasks such as image and speech recognition\citep{li2023deep, mukhamadiyev2022automatic}, natural language processing\citep{lauriola2022introduction}, and many other applications. A Similar application of the deep learning techniques could also be put to practical use for biological sequences\citep{jurtz2017introduction}.

Some of deep learning algorithms applied to protein sequences classification are listed in this section.

\subsubsection{Deep Neural Network (DNN)}
A conventional Deep Neural Network (DNN) is an expansion of the regular neural network (NN) model, introducing multiple hidden layers, each equipped with nonlinear structure to acquire layer-specific representations. Xiuquan Du et al. applied this technique to  learn the representations of proteins from common protein descriptors and protein-protein interactions (PPIs)\citep{du2017deepppi}. While this architecture has found widespread use across various applications, however, the training process tends to be sluggish and ineffective. The training of deep neural networks encounters challenges in terms of computational demands and time constraints influencing the feasibility and scalability of deep learning techniques in real-world applications. Consequently, there is a growing need for methods to reduce the complexities associated with training deep neural networks, which can significantly enhance their efficiency and applicability across a spectrum of machine learning tasks.

\subsubsection{The Recurrent Neural Network (RNN)}
The Recurrent Neural Network (RNN) model, an evolution from standard neural networks, is designed for discovering hidden patterns in sequential and temporal data streams. Unlike feedforward neural networks, which perform computations in a one-way flow from input to output, RNNs calculate the current state's output based on the outputs from previous states. However, this approach suffers from vanishing gradients over lengthy sequences. To overcome this limitation, two RNN variants, namely Long Short-Term Memory (LSTM) and Gated Recurrent Unit (GRU), were introduced. In the context of protein sequence analysis, the conventional approach involved comparing common phylogenetic motifs with known proteins, requiring significant domain experience and expert involvement. However, Nguyen Quoc Khanh Le et al.\citep{khanh2019classification} proposed an approach by incorporating recurrent neural networks (RNNs) and PSSM profiles to the classification of adaptor proteins. 

\subsubsection{The Convolutional Neural Network (CNN)}
The Convolutional Neural Network (CNN) is a multi-layer neural network model getting inspiration from the function of the visual cortex in neurobiology. It comprises convolutional layers followed by fully connected layers, with the inclusion of pooling functions to condense features between these types of layers. While originally devised for image processing\citep{liang2016cnn}, CNNs have obtained widespread acceptance in computer vision \citep{bhatt2021cnn} for their exceptional proficiency in detecting complex image patterns. This remarkable feature extends to the analysis of sequential data, making CNNs a versatile tool in diverse domains. As an example, Kim's work has demonstrated the adaptability of CNN networks in handling various sequential tasks, such as sentence-based classification\citep{kim2014convolutional}.  Additionally, Chaturvedi and colleagues applied pre-trained CNNs on network patterns derived from words and concepts, extracted from text via dynamic Gaussian Bayesian networks\citep{chaturvedi2016bayesian}. 

\subsubsection{The Long Short-Term Memory (LSTM)}
As previously mentioned, RNNs face difficulties when it comes to learning long-term dependencies in data sequences. LSTM-based models serve as an extension to conventional RNNs and effectively handle the vanishing gradient problem by incorporating a unique memory component referred to as a ``$gated$" cell. Within an LSTM model, essential features from the input data are captured and retained over prolonged periods. The decision to retain or discard this information is determined by the weight values assigned during the training phase. This characteristic renders LSTMs especially advantageous for tasks where maintaining context and managing extended dependencies is critical. Consequently, LSTMs excel in handling sequential data and prove exceptionally valuable for tasks involving sequences, such as those encountered in natural language processing\citep{lei2018sentiment} and time series analysis \citep{siami2019performance}.

\subsubsection{Bidirectional LSTMs (BiLSTM)}
BiLSTM, an extension of the LSTM architecture, strengthens its capabilities by processing input data bidirectionally, incorporating both forward and backward passes. This dual-directional strategy empowers the network to consider not only past context but also future context for each data point. This unique feature proves highly advantageous in various tasks, including sequence labeling\citep{pogiatzis2020using}, machine translation\citep{bensalah2021lstm}, and sentiment analysis\citep{yang2017attention}. This two-pass approach through the LSTM yields significant improvements in learning long-term dependencies within the data, ultimately leading to enhanced model accuracy. By considering both past and future information for each data point, BiLSTM excels in capturing comprehensive contextual information, making it a valuable asset in tasks that rely on understanding sequences and their inherent dependencies\citep{siami2019performance}

\begin{figure*}[t]
	\centering
	\includegraphics[width=0.80\textwidth]{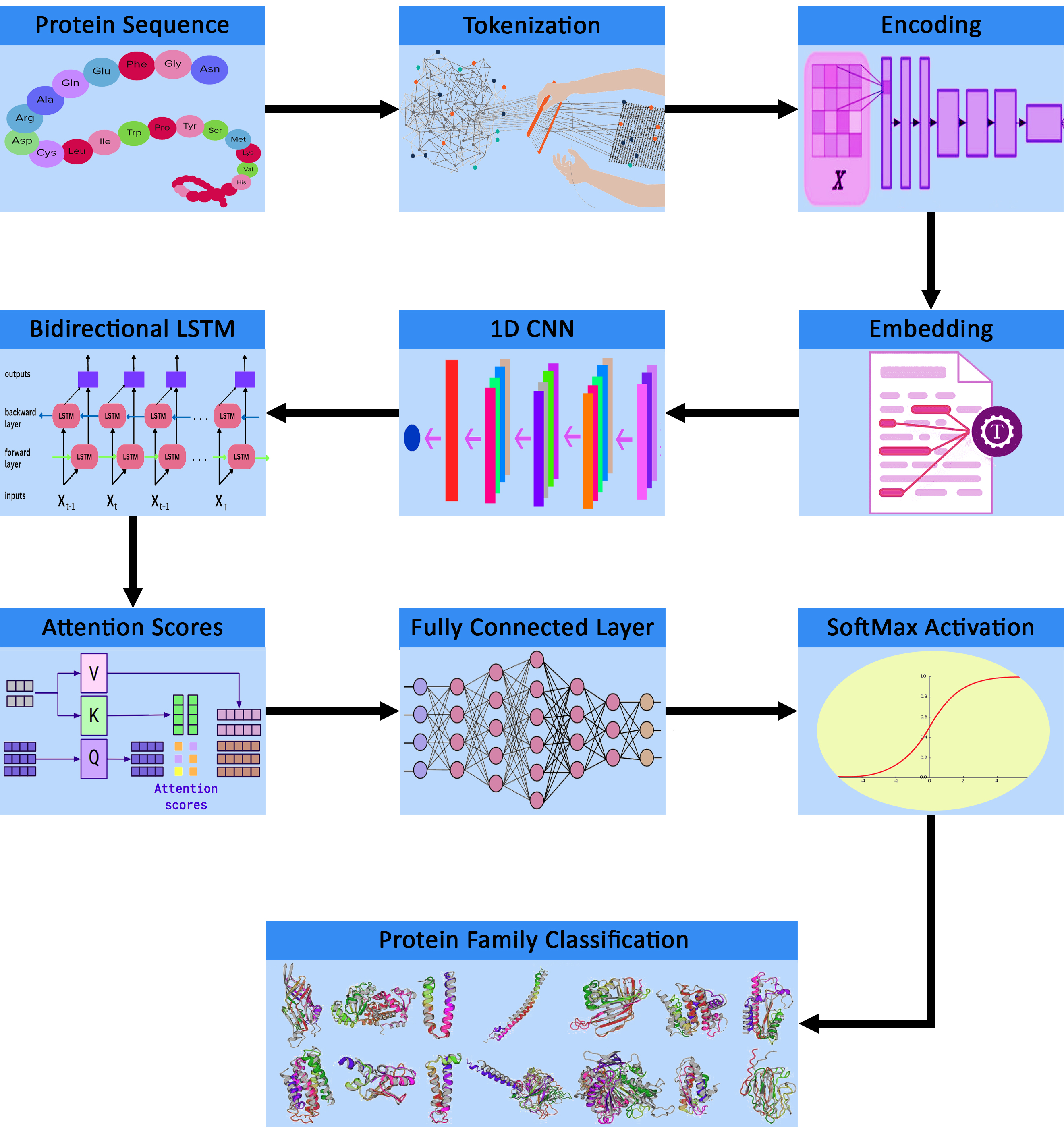}
	\caption{Visualizing ProFamNet: A graphical overview of layer structures}
	\label{fig:Architecture-Final}
\end{figure*}

\subsubsection{Attention mechanism}
The introduction of the attention mechanism has significantly enhanced the applicability of deep learning models across various domains. This mechanism effectively tackles the challenge of handling extensive long-range dependencies, a problem often encountered in LSTM-based models. As input sentences grow in length, traditional LSTMs face a decrease in performance as they struggle to retain connections between words that are widely separated within a sentence \citep{cho2014learning}. Notably, the pioneering work on attention mechanisms was initiated by Bahdanau and colleagues \citep{bahdanau2014neural}.

\begin{figure*}[t]
	\centering
	\includegraphics[width=0.99\textwidth]{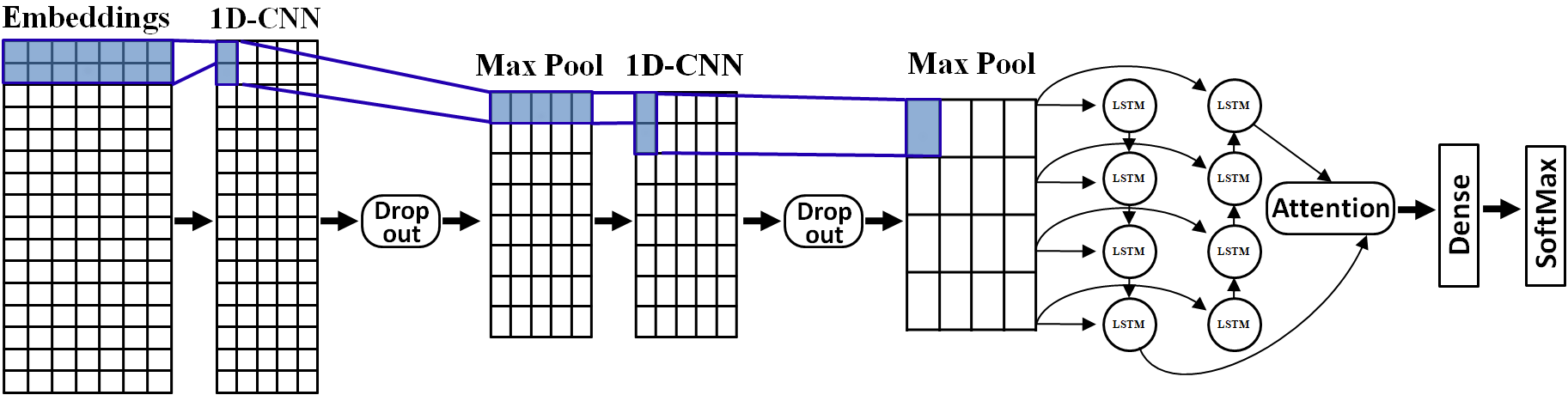}
	\caption{ProFamNet layer analysis: A structural overview}
	\label{fig:System_Architecture_1DCBLA}
\end{figure*}

\section{Proposed Work}
In this section, we introduce the novel deep learning approach aimed at enhancing and expediting the classification of protein functions, thereby delivering superior results. The proposed framework (ProFamNet) encompasses a series of modules applied sequentially, comprising encoding, embedding, 1D-CNN, and BiLSTM with an Attention mechanism. Each of these modules will be discussed in greater detail in the subsequent subsections. A Graphical Overview of Layer Structures the proposed model (ProFamNet) is shown in the Figure \ref{fig:Architecture-Final}. Further, A structural overview is shown in the Figure \ref{fig:System_Architecture_1DCBLA} depicting conversion process from one layer to the next layer.

\begin{table*}[t]
	\centering
	\begin{tabular}{||>{\raggedright}p{2.6cm}|c|c|c||}
		\hline
		\textbf{Amino Acid} & \textbf{Abbreviation} & \textbf{1 Letter} & \textbf{Quantization} \\
		\hline
		\hline
		Leucine & Leu & L & 1 \\
		\hline
		Alanine & Ala & A & 2 \\
		\hline   
		Glycine &Gly& G & 3 \\
		\hline
		Valine &Val& V & 4 \\
		\hline
		Glutamic acid &Glu& E & 5 \\
		\hline
		Isoleucine &Ile& I & 6 \\
		\hline
		Lysine& Lys &K& 7 \\
		\hline
		Arginine &Arg& R & 8 \\
		\hline
		Serine &Ser& S & 9 \\
		\hline
		Aspartic acid &Asp& D & 10 \\
		\hline
		Threonine &Thr& T & 11 \\
		\hline
		Proline &Pro& P & 12 \\
		\hline
		Phenylalanine &Phe& F & 13 \\
		\hline
		Asparagine &Asn& N & 14 \\
		\hline
		Glutamine &Gln & Q & 15 \\
		\hline
		Tyrosine &Tyr & Y & 16 \\
		\hline
		Methionine &Met & M & 17 \\
		\hline
		Histidine &His & H & 18 \\
		\hline
		Cysteine &Cys & C & 19 \\
		\hline
		Tryptophan &Trp & W & 20 \\
		\hline
		Any amino acid &Xaa & X & 21 \\
		\hline
		Aspartic acid or Asparagine &Asx & B & 22 \\
		\hline
		Glutamic acid or Glutamine&Glx & Z & 23 \\
		\hline
		Selenocysteine &Sec & U & 24 \\
		\hline
	\end{tabular}
	\caption{Encoding of amino acids }
	\label{tab:encoding_table}
\end{table*}

\subsection{Encoding Module} 
To begin with, ProFamNet requires the quantization of all the amino acids into a range of numbers. This process involves defining an alphabet set of size `$T$' to represent the input language and subsequently employing one-hot encoding to quantize each amino acid. Following this quantization, the encoding module comes into play, which is responsible for converting and transforming the sequence of amino acids into an integer array. In order to ensure the generality of the proposed approach across various scenarios, we have assigned unique integer values to each amino acid and have represented them as ${24}$ distinct numbers. These ${24}$ amino acids include: $L$, $A$, $G$, $V$, $E$, $I$, $K$, $R$, $S$, $D$, $T$, $P$, $F$, $N$, $Q$, $Y$, $M$, $H$, $C$, $W$, $X$, $B$, $Z$, $U$. The specific integer values corresponding to each amino acid are detailed in Table \ref{tab:encoding_table}.

For the purpose of analysis, we have restricted the utilization of protein sequences to those falling within a specific length range, specifically between 50 and 1200 amino acids. This selection aligns with the length distribution of protein sequences found in the UniProt\footnote{https://www.ebi.ac.uk/uniprot/TrEMBLstats} database. Sequences falling outside of this defined range have been excluded. This decision is partly motivated by the need to facilitate a fair comparison with previous research, where the same length criteria were applied to filter the sequences for consistency and benchmarking purposes.

\begin{figure*}[!]
	\centering
	\includegraphics[width=0.65\textwidth]{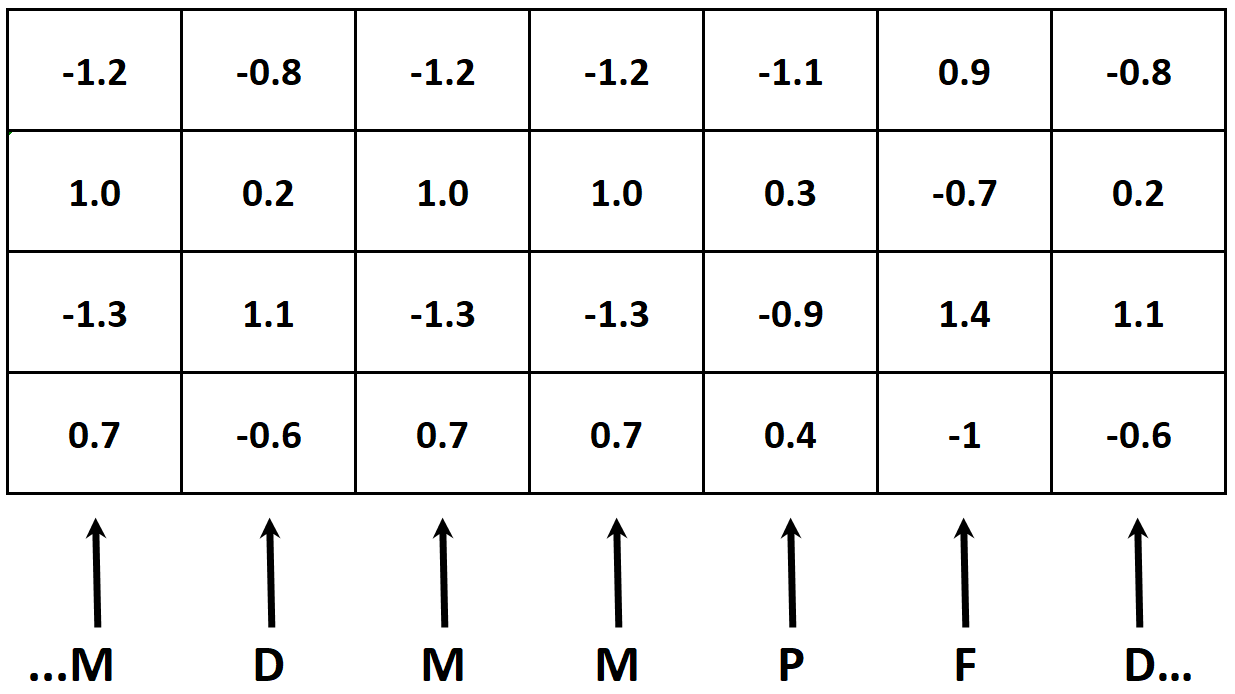}
	\caption{Embedding of partial protein sequence ...MDMMPFD...}
	\label{fig:Embedding_Partial_Protein_Sequence}
\end{figure*}

\subsection{Embedding Module} 
Following encoding, the embedding layer transforms each quantized amino acid value into a continuous vector. We follow Da Zhang et al\citep{zhang2020protein}, to translates each resultant number into a continuous vector within a fixed-dimensional space denoted as ${'W'}$. This enables continuous metric measures of similarity, thereby facilitating the assessment of the semantic attributes of individual amino acids. Following the embedding phase, we present a generated protein sequence matrix as an example, as illustrated in Figure \ref{fig:Embedding_Partial_Protein_Sequence}. A key advantage of ProFamNet method lies in the fact that the embedding process necessitates training only once for a predetermined number of epochs, after which it can be employed for encoding biological sequences, even for previously unseen protein sequences. This efficiency in the embedding process streamlines the application of ProFamNet approach in handling diverse biological sequences.  

\subsection{1D CNN} 
The temporal convolutional CNN module is responsible for performing 1D convolutions. Following the embedding phase, each protein sequence undergoes a transformation into a matrix. With this input matrix, the CNN module establishes a local connectivity pattern among the neurons in various layers, allowing for the exploitation of spatially local structures. More specifically, the CNN layers play a pivotal role in capturing non-linear features within protein sequences, uncovering motifs, and reinforcing higher-level associations.

It's worth noting that the mechanism for 1D convolution draws inspiration from the NLP domain, originally designed for text classification tasks. In the context of ProFamNet method, the matrix resulting from the embedding process is input to the CNN module, where feature detectors of different sizes (denoted as $'k'$) are employed to convolve over the entire matrix. It's essential to highlight that each feature detector consistently spans the full vector length.
For reference, the notations for the corresponding parameters utilized in ProFamNet can be found in table \ref{tab:mathematical_notations}.

\begin{table*}[t]
  \centering
  \begin{tabular}{||c|c||}
    \hline
    \textbf{Parameters} & \textbf{Notations} \\
    \hline
    \hline
    Total length of the protein sequence  & $\mathcal{L}$ \\
    \hline
    Embedding Vector length of each amino acid & $\mathcal{W}$ \\
    \hline
    Kernel Size & k \\
    \hline
    Convolutional filter matrix & $\mathbf{H}$ \\
    \hline
    Strid size & $s$ \\
    \hline
    Size of feature maps  & $\mathcal{M} = (\mathcal{L} - k) / s + 1$\\
    \hline
    Matrix representation of protein sequence & $\mathbf{C}$ \\
    \hline
  \end{tabular}
  \caption{List of mathematical notations}
  \label{tab:mathematical_notations}
\end{table*}

Referring to the table \ref{tab:mathematical_notations}, let's assume that, following the embedding module, we obtain an embedding input matrix denoted as $\mathbf{C} \rightarrow \mathbb{R} ^ {\mathcal{W} \times \mathcal{L}}$. In this context, we employ a filter (or kernel), $\mathbf{H} \in \mathbb{R} ^ {\mathcal{W} \times k}$, with a kernel size of \(k\), which is convolved over the input matrix. In this study, we employ the rectifier function represented as \(h(x) = \max(0, x)\) (commonly known as $\textbf{ReLUs}$) as the activation function to yield a feature map, \(f\), residing in the space of \(f \in \mathbb{R}^{\mathcal{H}-k+1}\). This feature map, \(\textbf{f}\), effectively serves as the output vector array. Consequently, the calculation of the \(i^{th}\) element within \(\textbf{f}\) can be determined using the equation \ref{eq:ithOfF}.

\begin{equation}
	f[i]=\left\langle \textbf{C} [*, i: i + k - 1], \textbf{H}\right\rangle 
	\label{eq:ithOfF}
\end{equation}

\begin{equation}
	y=\max_{i} ReLU(f[i]+b)
	\label{eq:ithOfF2}
\end{equation}

In this context, ${C[i: i + k - 1]}$ denotes the $i^{th}$ to ${(i + k - 1)^{th}}$ column of the input matrix $\textbf{C}$ and ${\left\langle .,. \right\rangle}$ representing inner product operations. A crucial element that facilitates the training of deeper models is the temporal max-pooling mechanism, initially derived from the field of computer vision. The specific max-pooling layer described in equation \ref{eq:ithOfF2} empowers us to train ConvNets deeper than six layers, an achievement unattainable by other methods \citep{boureau2010theoretical}. Consequently, this max-pooling mechanism allows us to perform a maximum operation over time on the output vector $f$. We interpret this resulting output $y$ as the feature corresponding to the filter $\textbf{H}$ for a given protein sequence $\textbf{C}$. Taking into account the stride size \(s\), the convolution \(y_j\) between each kernel function \(h(x)\) and the input function \(c(x)\) can be mathematically represented as in equation \ref{eq:convYOfI}.

\begin{equation}
	y(j) = \sum_{x=1}^{k} h(x) \cdot c(j \cdot s - x + k - s + 1)
	\label{eq:convYOfI}
\end{equation}

Where \(k-s + 1\) is an offset constant. If we have \(h\) filters in total (\(H_1, H_2, ..., H_h\)) and an input matrix \(C_j\), the output \(y^j = (y^j_1, y^j_2, ..., y^j_h)\) serves as the representation for the input matrix \(C_j\). In many NLP applications, \(h\) is typically chosen within the interval [100, 1000] \citep{kim2016character}.

Following the training process, the CNN module is characterized by a set of kernels \(\textbf{H}_i\), which we refer to as \textit{weights}, operating on a set of inputs \(\textbf{C}_k\). The outputs \(f_j\) are obtained by summing over \(i\) of the convolutions between \(\textbf{H}_i\) and \(\textbf{C}_k\). Consequently, with a kernel size of \(k\), the total number of iterations will be \((\mathcal{H} - k) / s+1\) with a stride size of \(s\) when convolving over the protein sequence matrix. This yields the output vector \(\textbf{f}\), having the same shape as \(o((\mathcal{H} - k) / s+1 \times 1)\). For instance, if we have a predefined protein sequence with a length of \(\textbf{H} = 1200\), a stride \(s = 1\), and a kernel size \(k = 3\), the total length of feature maps \(\mathcal{M}\) will be \((1200 - 3) / 1+1 = 1198\) after convolution. Unlike images with 2-dimensional (2D) pixel values, protein sequences exhibit 1D structures where each amino acid is significant. Therefore, for each filter, we fix one dimension as \(\mathcal{W}\) and vary the kernel size \(k\). The rationale behind the kernel size \(k\) is that a CNN with a kernel width \(k\) detects the presence of n-grams of length \(k\). For instance, with a kernel size of \(3\), the deep network can uncover \textit{tri}-gram information from the protein sequence data. Moreover, these \textit{tri}-grams can also be considered as motif representations for these protein sequences. Consequently, in this 1D-CNN framework, we can interpret the kernel size as the number of amino acid rows, indicating the number of amino acids within the protein matrix to be filtered. Each kernel is employed to convolve through the sequence data to capture specific features. After the data convolution, we obtain the feature map \(\textbf{f}\), add a bias term, and apply an activation function (ReLU in equation \ref{eq:ithOfF2}). We employ a max-pooling layer between the convolutional layers to select the optimal feature by taking the maximum value from multiple features. The above process yields one feature from one filter matrix. This approach employs multiple filters of varying widths to obtain feature vectors.

In summary, after embedding a protein sequence into \(X = [x_1 \cdots x_t \cdots x_T]\), where \(X \in \mathbb{R}^{D \times T}\), with \(D\) being the dimension of \(x_t\, \) we aim to obtain an abstract-level sequence representation \(Z = [z_1 \cdots z_t \cdots z_T]\), where \(Z \in \mathbb{R}^{D_{\text{out}} \times T}\) and \(D_{\text{out}}\) is the dimension of \(z_t\), achieved through 1D convolutions applied over the sequential dimension of the sequence \(X\).

\begin{equation}
	Z = \mathcal{S}(X)
	\label{eq:dimension_Z} 
\end{equation}

In equation \ref{eq:dimension_Z} $\mathcal{S}$ represents a stacked 1D fully convolutional neural network. It leverages convolution operations to extract fixed-size contextual information from all data points within its receptive field. Consequently, when we input the sequence X into the deep 1D-CNN framework, the resulting output Z has already integrated both the spatial and temporal characteristics of the sequence through its expansive receptive field. This renders the new sequence Z sufficiently distinctive for classification.


\subsection{LSTM and BiLSTM Module} 
LSTMs, short for Long Short-Term Memory networks, are an extension of RNNs. LSTMs introduce specialized hidden units designed to preserve memory . LSTMs are well-suited to effectively mitigate the vanishing gradient problem that commonly occurs in traditional RNNs. Each repeating unit within an LSTM consists of the essential neural components: the forget gate layer, the input gate layer, which comprises an update component and an addition component, and the output gate layer. These components interact with the cell state, a distinct memory cell that accumulates and transmits information throughout the model. These interactions occur through gates that regulate the addition and modification of data in the cell state.

Specifically, the forget gate layer decides the amount of information to discard or remove. The input gate layer manages the integration of values requiring updates and potential additions. Finally, the output gate layer governs which parts of the internal memory state should be presented as output \citep{lecun2015deep, liu2016recurrent}. The LSTM's ability to retain information in its memory makes it highly suitable for handling sequential data, such as protein sequences.

Let's examine an LSTM unit at a particular time step, denoted as $t$. Within this context, $in_t$ takes on the role of the input gate, $f_t$ operates as the forget gate, and $op_t$ represents the output gate. The internal cell state is symbolized as $c_t$, while $o_t$ stands for the unit's output, often referred to as the hidden state. The operation of an LSTM is guided by the set of equations \ref{eq:lstm_in_t} through \ref{eq:o_t}:

\begin{equation}
	in_t = \sigma(W_{ip} \cdot i_t + U_{ip} \cdot o_{t-1} + b_{ip})
	\label{eq:lstm_in_t} 
\end{equation}

\begin{equation}
	f_t = \sigma(W_f \cdot i_t + U_f \cdot o_{t-1} + b_f)
	\label{eq:lstm_f_t} 
\end{equation}

\begin{equation}
	op_t = \sigma(W_{op} \cdot i_t + U_{op} \cdot o_{t-1} + b_{op})
	\label{eq:lstm_op_t} 
\end{equation}

\begin{equation}
	\tilde{c}_t = \tanh(W_c \cdot i_t + U_c \cdot o_{t-1} + b_c)
	\label{eq:lstm_tct} 
\end{equation}

\begin{equation}
	c_t = f_t \odot c_{t-1} + in_t \odot \tilde{c}_t
	\label{eq:lstm_ct} 
\end{equation}

\begin{equation}
	o_t = op_t \odot \tanh(c_t)
	\label{eq:o_t} 
\end{equation}

In these equations, the symbol $\sigma$ signifies the logistic sigmoid function and is responsible for element-wise multiplication. The $W$ and $U$ vectors denote the respective weight parameters, while $b$ represents the bias terms. The term $\tilde{c}_t$ signifies a vector of candidate values for the memory cell state, from which the actual cell state $c_t$ is derived by merging it with the input gate.

Equation \ref{eq:lstm_in_t} represents the process of assessing both the previous output $o_{t-1}$ and the current input $i_t$ through a sigmoid layer to determine the relevant information to include, essentially acting as the input gate. 

Equation \ref{eq:lstm_f_t} characterizes the forget gate, which employs a similar mechanism to selectively discard specific information. Finally, equation \ref{eq:lstm_op_t} defines the output operation. The model incorporates a sigmoid layer designed to make decisions about which states to propagate. Equation \ref{eq:lstm_tct} characterizes the current moment information $\tilde{c}_t$, obtained by applying a hyperbolic tangent (\text{tanh}) function to the combination of the preceding output and the current input. This newly generated information is slated for addition to the overall computation. Importantly, each of these gates relies on its specific weight vectors ($W$ and $U$) and bias vectors ($b$) to contribute to the overall decision-making process, emphasizing the role of these parameters in shaping the model's behavior and learning dynamics.
In equation \ref{eq:lstm_ct}, the computation involves the multiplication of the current memory information, obtained from the output of equations \ref{eq:lstm_in_t} and \ref{eq:lstm_tct}, with the input gate information. Simultaneously, this result is added to the product of the forget gate information and the long-term memory information from the previous time step, denoted as $c_{t-1}$. This  process leads to the generation of the long-term memory information at the current time step ($c_t$). Subsequently, this information undergoes transformation through a hyperbolic tangent (\text{tanh}) layer. The resulting output is further multiplied by the outcome of the output gate computation in equation \ref{eq:lstm_op_t}, ultimately yielding the final output of the Long Short-Term Memory (LSTM) model, denoted as $o_t$.

Utilizing Bidirectional Long Short-Term Memory networks (BiLSTMs) involves applying LSTM to the input data in both forward and backward directions. The results of employing this approach consistently show that the supplementary training offered by BiLSTMs leads to enhanced predictivity.

\subsection{Attention Mechanism} 
\begin{figure}
  \centering
  \includegraphics[width=0.45\textwidth]{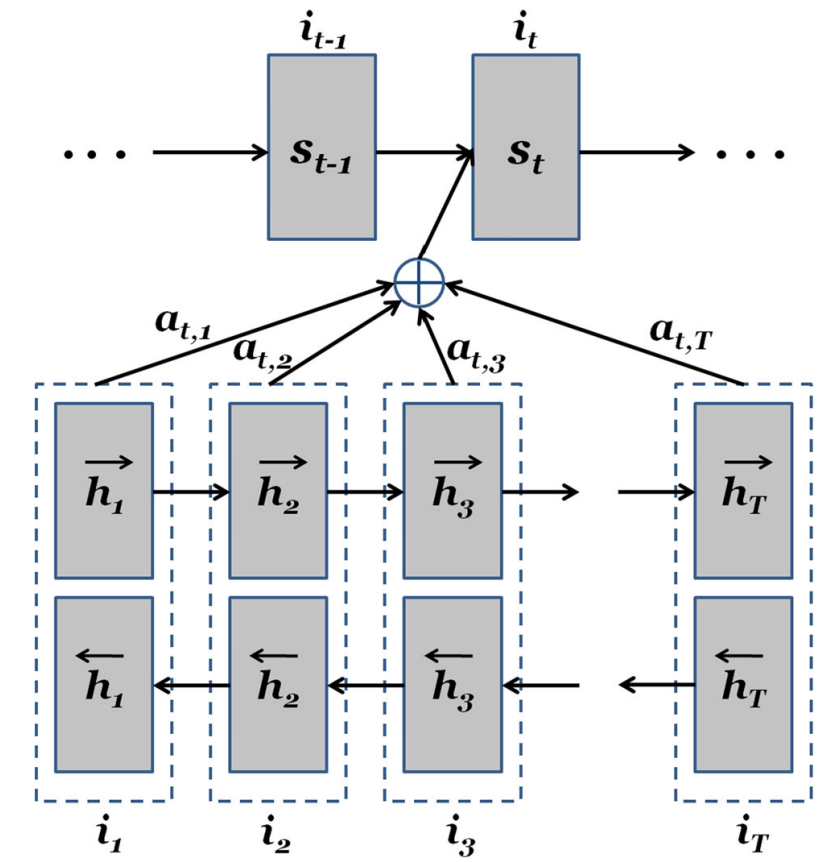}
  \caption{Attention illustration by Bahdanau et al. \citep{bahdanau2014neural}}
  \label{fig:Attention_illustration}
\end{figure}

The first attention model, proposed by Bahdanau et al. \citep{bahdanau2014neural} and depicted in Figure \ref{fig:Attention_illustration}, has exceedingly influenced diverse domains, particularly in natural language processing (NLP). This mechanism addresses very long-range dependency challenges in Long Short-Term Memory networks (LSTMs). As input sentence length increases, LSTM performance decreases due to reduced ability to retain connections between distant words \citep{cho2014learning}. Additionally, LSTMs lack the inherent capacity to prioritize relevant sentence sections. The attention mechanism efficiently addresses both issues by creating context vectors, taking into account all input words, and assigning relative weights based on their significance. This breakthrough, initiated by Bahdanau et al.\citep{bahdanau2014neural}, has significantly enhanced deep learning models, particularly in NLP, allowing for improved comprehension of extensive textual data and marking a crucial advancement in neural network capabilities. The Figure \ref{fig:Attention_illustration} illustrates that the model utilizes a bidirectional RNN as an encoder, enabling it to analyze the input sentence in both forward and backward directions for a holistic comprehension of contextual connections among words. The encoder generates states, representing individual words, with each state encompassing information about the entire sentence while giving emphasis to specific phrases and their adjacent context. This adaptive approach allows the model to effectively monitor and focus attention on pivotal words across the sentence. The encoder states, initially starting with the last encoder state, are combined with the current decoder state to train a feed-forward neural network. This network generates unique scores, representing the attention given to each encoder state. By applying the softmax function to these scores, attention weights are produced. These weights are then used to compute a context vector for the input sentence, offering a dynamic representation that contrasts with the fixed context vector of traditional RNNs, which rely solely on the last form. 

The decoder, implemented as a gated recurrent unit (GRU), utilizes the dynamically generated context vector alongside the output from the preceding time step to predict the next word in the sentence. This predicted word then becomes the current state of the decoder. This iterative process continues as the current state is recurrently combined with the encoder states, refining a feed-forward neural network responsible for computing scores that signify the importance of each encoder state.

This cyclic interplay persists until the decoder generates an 'END' token, indicating the completion of the sentence generation. The cohesive synergy between the GRU-based decoder, the evolving context vector, and the recurrent refinement of the feed-forward neural network ensures the generation of a grammatically coherent and contextually meaningful sentence. \citep{lamba2019intuitive}

Let \(x = (x_1, x_2, \ldots, x_{Tx})\) denote an input sequence, where \(x_j\) represents an individual word. The annotation \(h_j\) is defined as given in the equation \ref{eq:backforward_concate},  
representing the concatenation of a forward hidden state \(\overrightarrow{h_j}\) and a backward hidden state \(\overleftarrow{h_j}\). The bidirectional Recurrent Neural Network (RNN) yields two sets of hidden states: left-to-right forward hidden states \((\overrightarrow{h_1}, \overrightarrow{h_2}, \ldots, \overrightarrow{h_{Tx}})\) and right-to-left backward hidden states \((\overleftarrow{h_1}, \overleftarrow{h_2}, \ldots, \overleftarrow{h_{Tx}})\). At each time step \(i\), the context vector is computed as the weighted sum of annotations, as specified by equation \ref{eq:attention_ci}.

\begin{equation}
	\overrightarrow{h_{j}^{T}}  ;  {\overleftarrow{h_{j}^{T}}}^T
	\label{eq:backforward_concate} 
\end{equation}

\begin{equation}
	c_i = \sum_{j=1}^{Tx} \alpha_{ij} h_j
	\label{eq:attention_ci} 
\end{equation}

The weight \(\alpha_{ij}\) assigned to each annotation \(h_j\) is computed using a softmax function as in equation \ref{eq:attention_alphaij}: 

\begin{equation}
	\alpha_{ij} = \frac{\exp(e_{ij})}{\sum_{k=1}^{Tx}\exp(e_{ik})}
	\label{eq:attention_alphaij} 
\end{equation}

This term \(\alpha_{ij}\) serves as a probabilistic measure, indicating the significance of annotation \(h_j\) in determining the subsequent hidden state \(s_i\) from the preceding hidden state \(s_{i-1}\). In essence, \(c_i\) represents the anticipated annotation, considering all annotations with probabilities \(\alpha_{ij}\).

The value \(e_{ij}\) given in equation \ref{eq:attention_eij} is derived as the output of an alignment model \(a()\), mapping the input at step \(j\) and the output at step \(i\). It functions as a score, generated by assessing the alignment between the $j$th annotation of the input (i.e., \(h_j\)) and the output of the preceding hidden state \(s_{i-1}\). This score captures the alignment quality, signifying the degree of correspondence between input annotations and the output hidden state at the current time step. The softmax function then normalizes these scores to obtain probability weights, crucial for determining the relevance and contribution of each annotation to the context vector \(c_i\).

\begin{equation}
	e_{ij} = a(s_{i-1}, h_j)
	\label{eq:attention_eij} 
\end{equation}

The alignment model functions as a feed-forward neural network, strategically designed to compute a soft alignment that facilitates the efficient back propagation of the gradient of the cost function. This alignment model plays a pivotal role in implementing the attention mechanism within the decoder. The underlying principle involves a probability calculation that distributes information across the annotation sequence. This strategic dissemination allows the decoder to selectively retrieve relevant information, contributing to the model's adaptability in handling diverse input sequences.

Initially conceived as an effective solution for machine translation tasks, the attention mechanism has evolved into a versatile tool for various text mining applications. Its inherent ability to selectively attend to relevant portions of the input sequence has proven beneficial across a spectrum of NLP tasks. Notably, the integration of attention mechanisms has demonstrated a substantial enhancement in the performance of models based on Long Short-Term Memory (LSTM) networks. Recognizing this improvement, the researchers have chosen to incorporate the attention mechanism into the ProFamNet model for the current study, taking advantage of its proven efficacy in capturing detailed relationships within sequential data.

\subsection{Fully connected module} 
Following the extraction of deep features through the utilization of various kernel sizes, these features are concatenated, flattened, and subsequently fed into a fully connected (FC) module for classification. This FC module encompasses multiple fully connected layers, culminating in an output layer that employs a softmax function to generate a probability distribution across all available classes. Consequently, the quantity of protein families dictates the number of units within this final output layer, forming a crucial architectural consideration.

Subsequently, the error derived from the final classification layer undergoes backpropagation throughout the network. In the course of the learning process, the loss function of the neural network is defined as the categorical cross-entropy between the predictions and the target values. Denoting the number of protein families as $C$, the network loss as $\mathcal{L}$, and the mini-batch size as $N$, the loss function is formally expressed in equation \ref{eq:FC_loss}. This framework ensures that the neural network optimally adapts its parameters to minimize the discrepancy between predicted and actual class assignments, essential for effective classification in the context of protein family classification.

\begin{equation}
	\mathcal{L} = -\frac{1}{N} \sum_{i=1}^{N} \sum_{c=1}^{C} I_{y_i \in C_c} \log p_{\textit{model}}[y_i \in C_c]
	\label{eq:FC_loss} 
\end{equation}

In this context, each observation \(i\), \(N\) in total are combined across  \(C\) categories individual category being represented by \(c\). The notation \(I_{y_i} \in C_c\) denotes the actual class \(C_c\) of observation \(y_i\), while \(p_{\textit{model}}[y_i \in C_c]\) specifies the model's predicted probability for \(y_i\) belonging to class \(C_c\). When dealing with multiple classes, the neural network produces a \(C\)-dimensional output vector, assigning probabilities to each class. Consequently, during training, our focus is on minimizing the loss function across all \(N\) observations, refining the model's predictive capabilities for future data.

This iterative process of minimizing the loss function not only optimizes the model's performance but also facilitates its generalization to new, unseen data. Strategic adjustments to model parameters enhance adaptability, ensuring reliable predictions in diverse real-world applications beyond the confines of the training dataset. This approach contributes to the neural network's overall robustness and efficacy, positioning it as a versatile tool for predictive modelling across various datasets and application domains.

\subsection{The overall system architecture}
\begin{figure*}[t]
	\centering
	\includegraphics[width=0.98\textwidth]{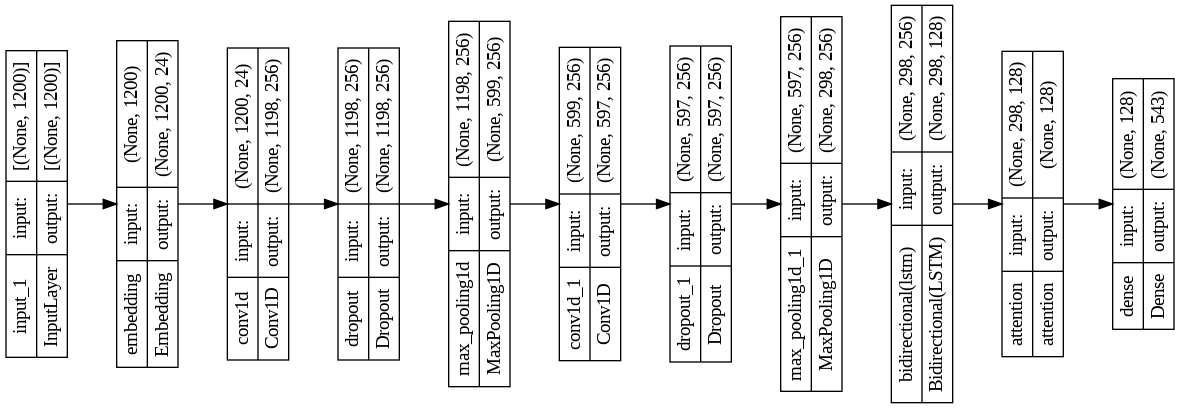}
	\caption{ProFamNet - model layers configuration}
	\label{fig:this_paper_sequence_diagram_horizontal}
\end{figure*}

ProFamNet is built with a depth of 5 layers, including 2 layers of 1D-CNN, 1 layer of BiLSTM, 1 layer of Attention, and 1 Dense layer. The complete layer configuration is illustrated in the Figure \ref{fig:this_paper_sequence_diagram_horizontal}. We have incorporated dropout layers before each 1D-CNN layer as a means of regularization, aiming to address the challenge of overfitting.

The overall architecture of the proposed system (ProFamNet) is given in the Figure \ref{fig:Architecture-Final}. The input to the system is a protein sequence, representing the primary structure of a protein. This sequence contains information about the amino acids that constitute the protein.Before feeding the sequence into the model, we perform preprocessing steps. These steps may involve data cleaning, normalization, and handling missing values. The protein sequence is tokenized into smaller units (e.g., amino acids or n-grams). Each token is encoded into a numerical representation suitable for neural networks. The encoded tokens are transformed into dense vector representations (embeddings). These embeddings capture semantic relationships between tokens. The 1D-CNN layer scans the sequence to detect local patterns and features. It learns filters that capture relevant motifs within the protein. The BiLSTM layer processes the sequence bidirectionally, capturing long-range dependencies and context information. Attention mechanisms highlight important regions of the sequence. These scores guide the model’s focus during decision-making. The fully connected layer aggregates information from previous layers, preparing the features for the final prediction. The SoftMax activation function produces probability scores for protein family classes, assigning likelihoods to different protein families. Based on the SoftMax output, the model predicts the protein’s family or functional category. This prediction aids in understanding the protein’s biological role.

\section{Experimental results and discussion}
\subsection{Evaluation Metrics}
In evaluating the efficacy of ProFamNet method, we employ a comprehensive set of metrics, including Accuracy, Recall, Precision, and F1-score. The definitions of these metrics are precisely outlined through equations \ref{eq:accuracy} to \ref{eq:f1score}, offering a formalized understanding of their calculations. Accuracy serves as a fundamental measure, reflecting overall correctness, while Recall assesses the model's ability to capture relevant instances within a class. Precision, focuses on the accuracy of positive predictions. The F1-score strikes a balance between Precision and Recall, providing a nuanced evaluation that considers both false positives and false negatives. This multifaceted approach ensures a thorough assessment of proposed method's performance, offering insights into various dimensions and contributing to a robust understanding of its strengths and potential areas for improvement.

\begin{equation}
\text{Accuracy} = \frac{(TP + TN)}{(TP + TN + FP + FN)}
\label{eq:accuracy}
\end{equation}

\begin{equation}
\text{Recall} = \frac{TP}{(TP + FN)}
\label{eq:recall}
\end{equation}

\begin{equation}
\text{Precision} = \frac{TP}{(TP + FP)}
\label{eq:precision}
\end{equation}

\begin{equation}
\text{F1-Score} = \frac{(2 \times \text{recall} \times \text{precision})}{(\text{recall} + \text{precision})}
\label{eq:f1score}
\end{equation}

\subsection{Experimental setup, datasets and results}
In the proposed work, the authors have utilized protein sequence data taken from the comprehensive UniProt\footnote{https://www.uniprot.org/uniprotkb/?query=reviewed=yes} database, encompassing all the reviewed entries.

In total, this database contains $271,160$ protein sequences of varying lengths across $543$ protein families. Initially we collected the entire UniProt database; subsequently, we applied filtering criteria, excluding sequences with lengths below $50$ and exceeding $1,200$. Further refinement involved the exclusion of protein families with fewer than $200$ instances, enabling a meaningful comparison with alternative methods. The process included the fine-tuning of hyper-parameters known as model configuration, encompassing variables such as the number of epochs, fully-connected dropout rate, learning rate, loss function, batch size, and optimizer as provided in Table \ref{tab:model_configuration}. 

Following careful tuning, ProFamNet model consistently employed the same set of hyperparameters outlined in Figure \ref{fig:this_paper_sequence_diagram_horizontal}, featuring $5$ layers in total. In this architecture, the initial $2$ 1D-CNN layers were designed to extract profound and abstract features from the input data. These features could subsequently be leveraged for downstream tasks, such as protein classification, through the application of BiLSTM, Attention and Fully Connected layers. Upon model training, predictions were made on test inputs, and receiver operating characteristic (ROC) curves were plotted for each class. Throughout the experiments, we maintained a dataset distribution ratio of $70\%$ for training, $15\%$ for validation, and $15\%$ for testing, ensuring a robust evaluation of the proposed model's performance.


\begin{table}[t]
\centering
\begin{tabular}{||l|l||}
\hline
\textbf{Optimizer} & Adam \\
\hline
\textbf{Learning Rate} & 0.001 \\
\hline
\textbf{Loss Function} & Categorical Cross Entropy \\
\hline
\textbf{Number of Epochs} & 25 \\
\hline
\textbf{Dropout rate} & 0.1 \\
\hline
\textbf{Batch Size} & 128 \\
\hline
\end{tabular}
\caption{Model configuration}
\label{tab:model_configuration}
\end{table}

\subsection{Performance comparisons with other methods}
We also conducted a performance comparison between the proposed system and other cutting-edge methods, utilizing diverse N-gram features on the UniProt dataset. The comprehensive results, presented in Table \ref{tab:model_comparison}, demonstrate that ProFamNet attains an impressive F1-score of $98.30\%$ on the validation dataset and performs exceptionally well with a score of $98.32\%$ on the testing dataset. In terms of F1-score, the proposed approach outperforms leading methods using the GRU model by $3\%$ on an unseen dataset and significantly outpaces the traditional SVM method utilizing tri-gram features by over $11\%$. The proposed approach also outperformed the latest Da Zhang et al.\citep{zhang2020protein}  by $1\%$. In Table \ref{tab:head_to_head_comparison}, we have presented a head to head comparison of the proposed work with the work of Da Zhang et al. \citep{zhang2020protein}.

\begin{table}[t]
\centering
\begin{tabular}{||>{\raggedright}p{2cm}|c|c|c||}
\hline
\textbf{Architecture} & \textbf{Dropout} & \textbf{Validation F1} & \textbf{Test F1} \\
\hline\hline
LSTM\citep{lee2016protein} & 0.85 & 92.61\% & 92.51\% \\
\hline
biLSTM\citep{lee2016protein} & 0.7 & 92.56\% & 92.22\% \\
\hline
GRU\citep{lee2016protein} & 0.8 & 95.31\% & 94.84\% \\
\hline
CNN (Three layers) \citep{lee2016protein} & 0.8 & 93.4\% & 93.4\% \\
\hline
SVM\citep{lee2016protein} & N/A & N/A & 87.88\% \\
\hline
CNN\citep{zhang2020protein}  & 0.6 & 97.67\% & 97.77\% \\
\hline
\textbf{ProFamNet} & \textbf{0.1} & \textbf{98.30\%} & \textbf{98.32\%} \\
\hline
\end{tabular}
\caption{Comparison of prediction performance across various models}
\label{tab:model_comparison}
\end{table}

\begin{table*}[t]
    \centering
    \begin{tabular}{||l|c|c||}
\hline
        & \textbf{Da Zhang et al.\citep{zhang2020protein}} & \textbf{ProFamNet} \\
\hline
\hline
        No. of Classes & 222 & 543 \\
\hline
		No. of parameters & 4,578,911 & 450,953 \\
\hline        
        Size & 17.47 MB & 1.72 MB \\
\hline
        No. of layers & 10 & 5 \\
\hline
        No. of epochs & 30 & 25 \\
\hline
        Validation F1 & 97.67\% & 98.30\% \\
\hline
        Test F1 & 97.77\% & 98.32\% \\
\hline
    \end{tabular}
    \caption{Head to head comparison with Da Zhang et al.\citep{zhang2020protein}}
    \label{tab:head_to_head_comparison}
\end{table*}
\thispagestyle{empty} 



\begin{figure*}[h]
	\centering
	\begin{subfigure}[b]{0.48\textwidth}
		\includegraphics[width=\textwidth]{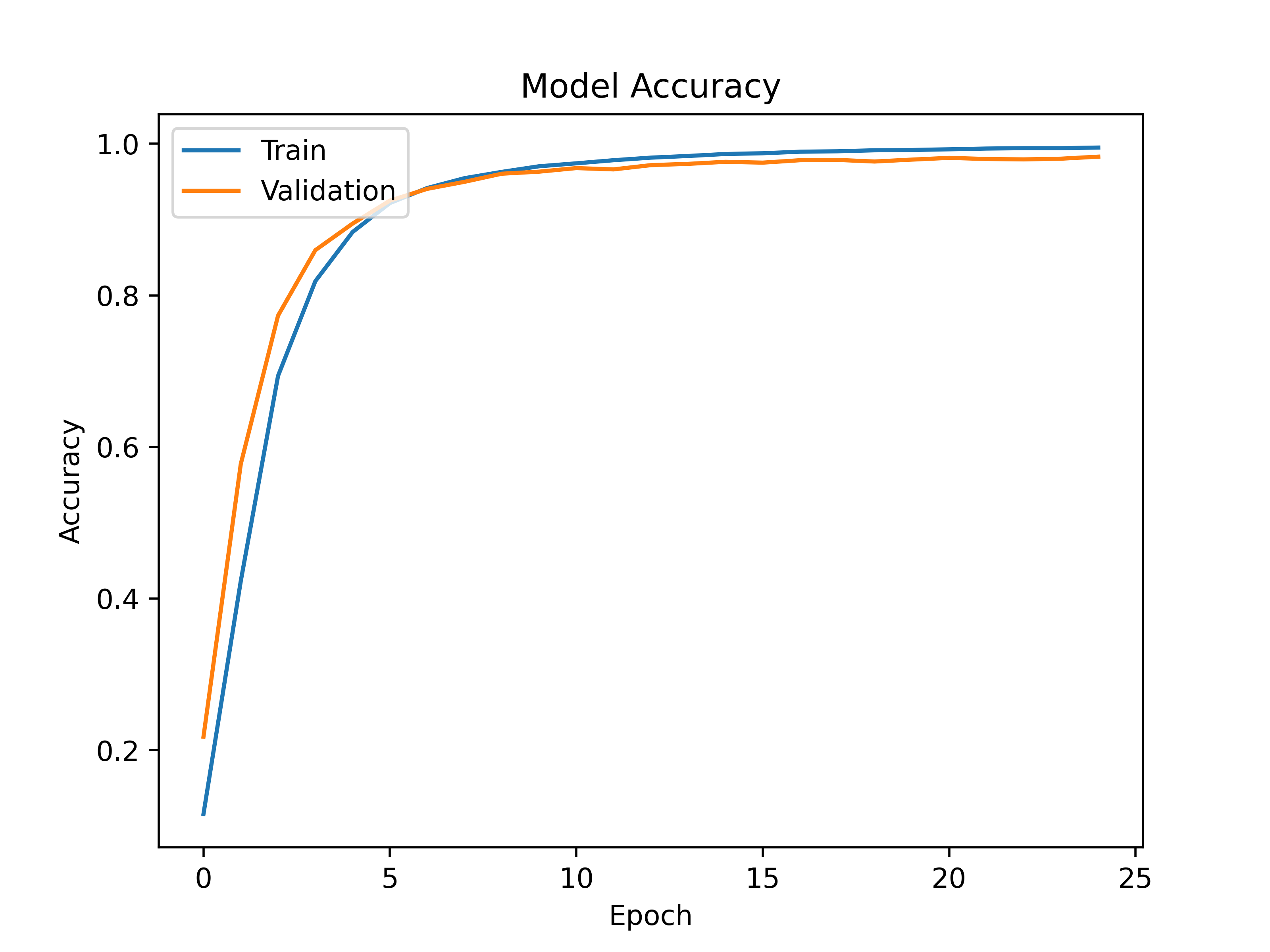}
		\caption{Accuracy of ProFamNet}
		\label{fig:our_model_accuracy}
	\end{subfigure}
	\begin{subfigure}[b]{0.48\textwidth}
		\includegraphics[width=\textwidth]{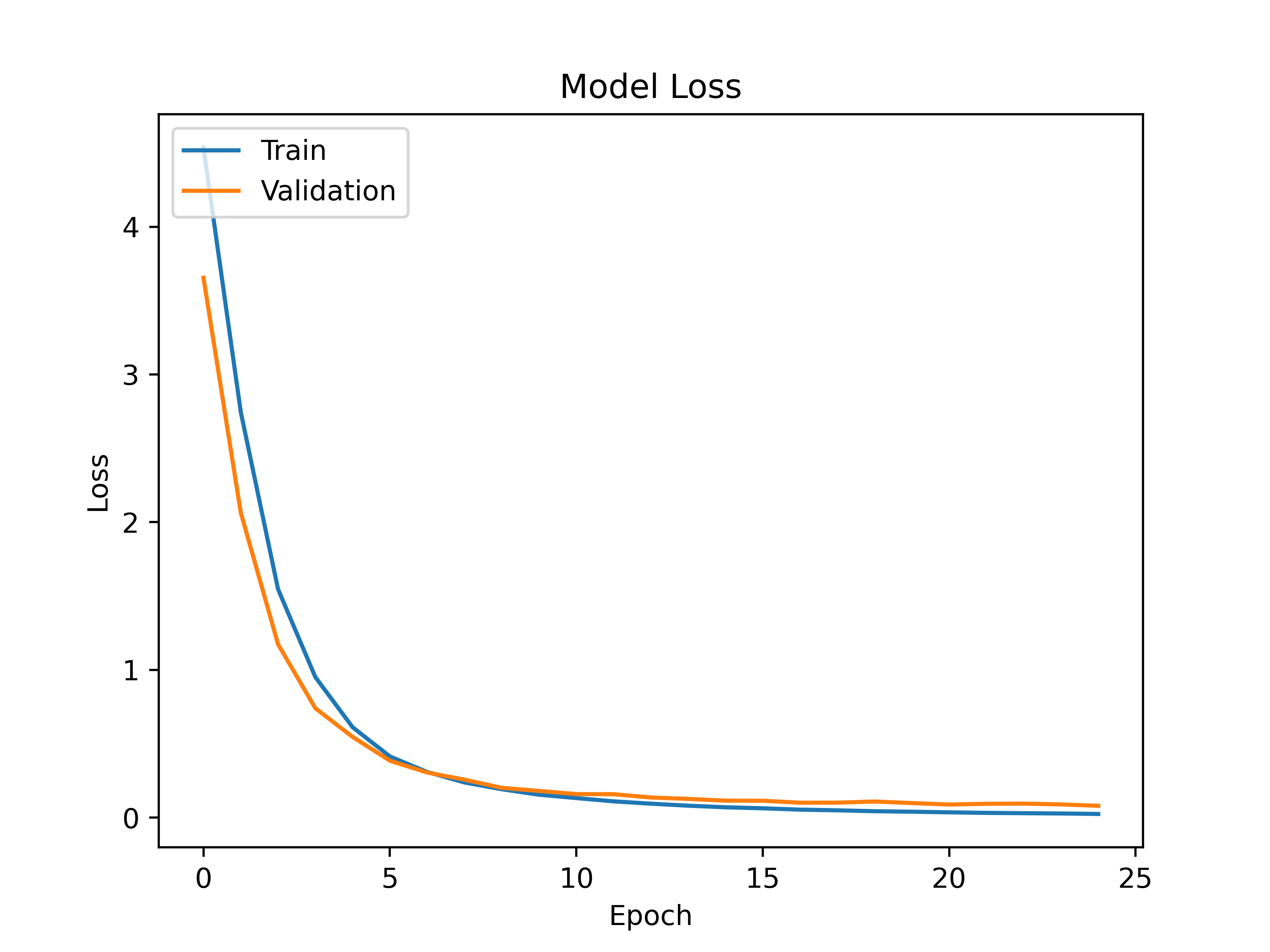}
		\caption{Loss of ProFamNet}
		\label{fig:our_model_loss}
	\end{subfigure}
	
	\caption{Accuracy \& loss of ProFamNet}
	\label{fig:our_model_accuracy_and_loss}
\end{figure*}

\begin{figure*}[t]
  \centering
  \begin{subfigure}[b]{0.48\textwidth}
    \includegraphics[width=\textwidth]{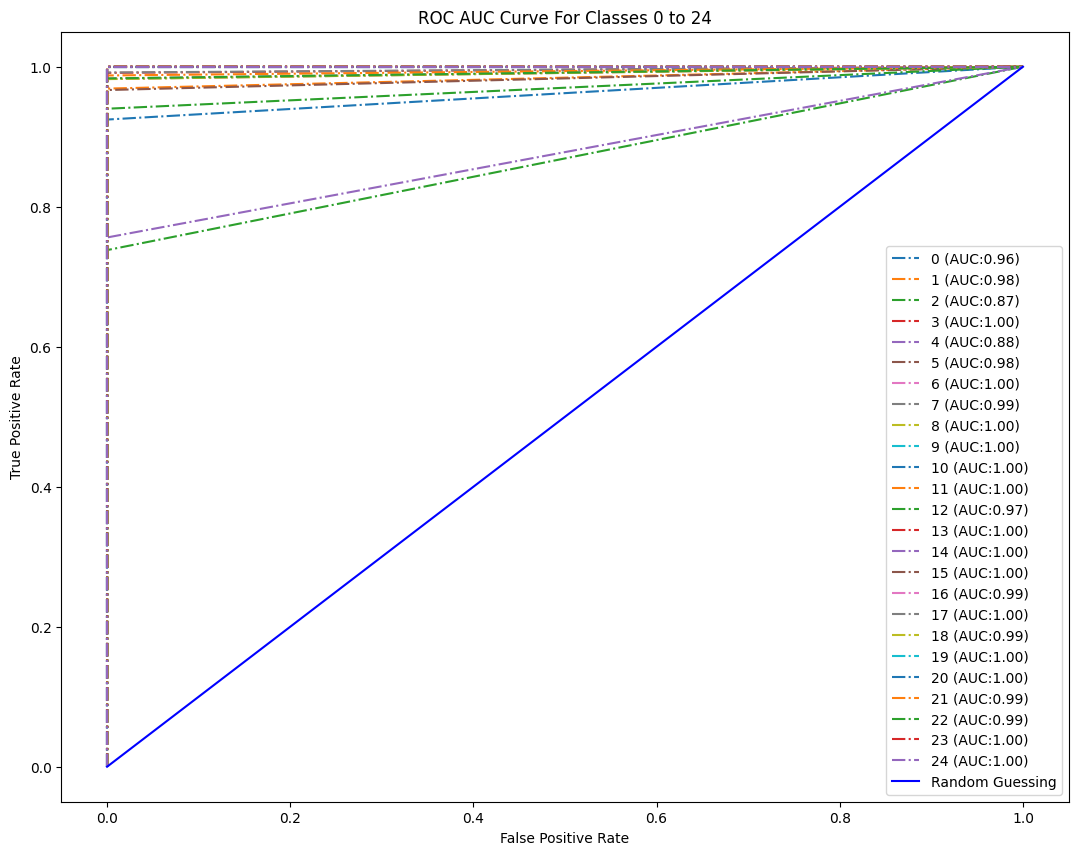}
    \caption{AUC-ROC scores 0-24}
  \end{subfigure}
  \begin{subfigure}[b]{0.48\textwidth}
    \includegraphics[width=\textwidth]{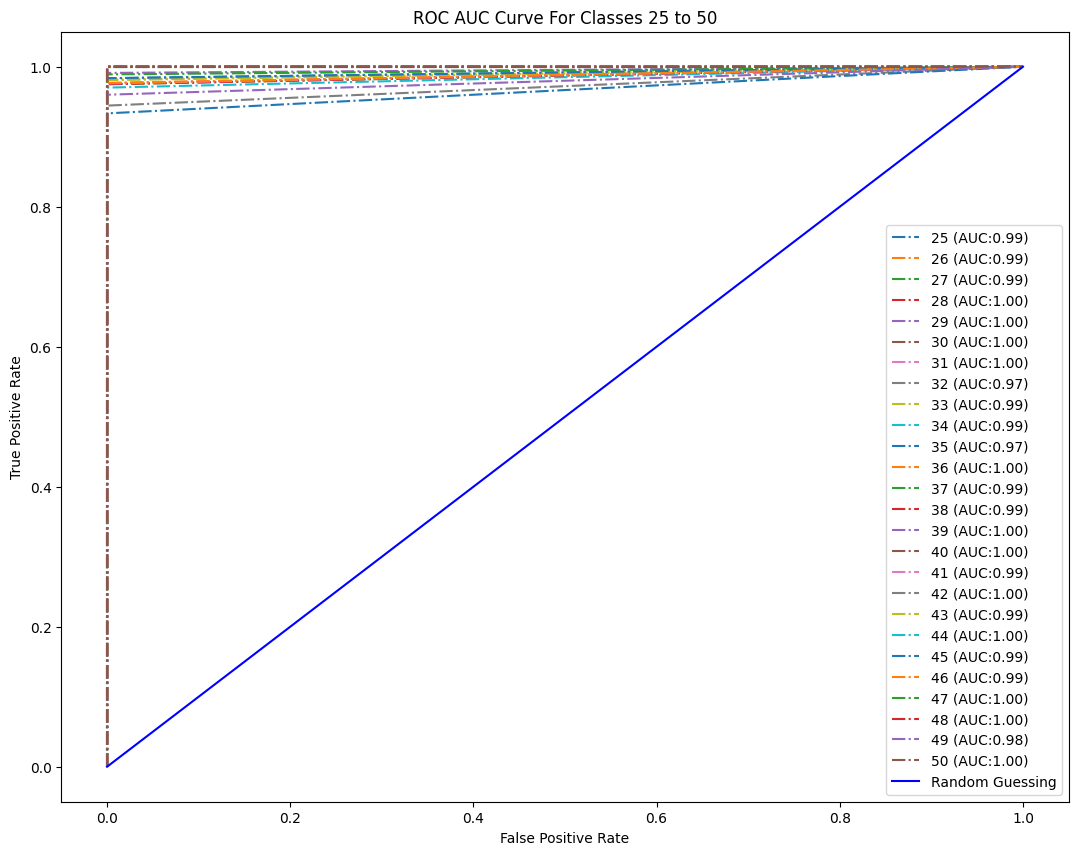}
    \caption{AUC-ROC scores 25-50}
  \end{subfigure}
  
  \begin{subfigure}[b]{0.48\textwidth}
    \includegraphics[width=\textwidth]{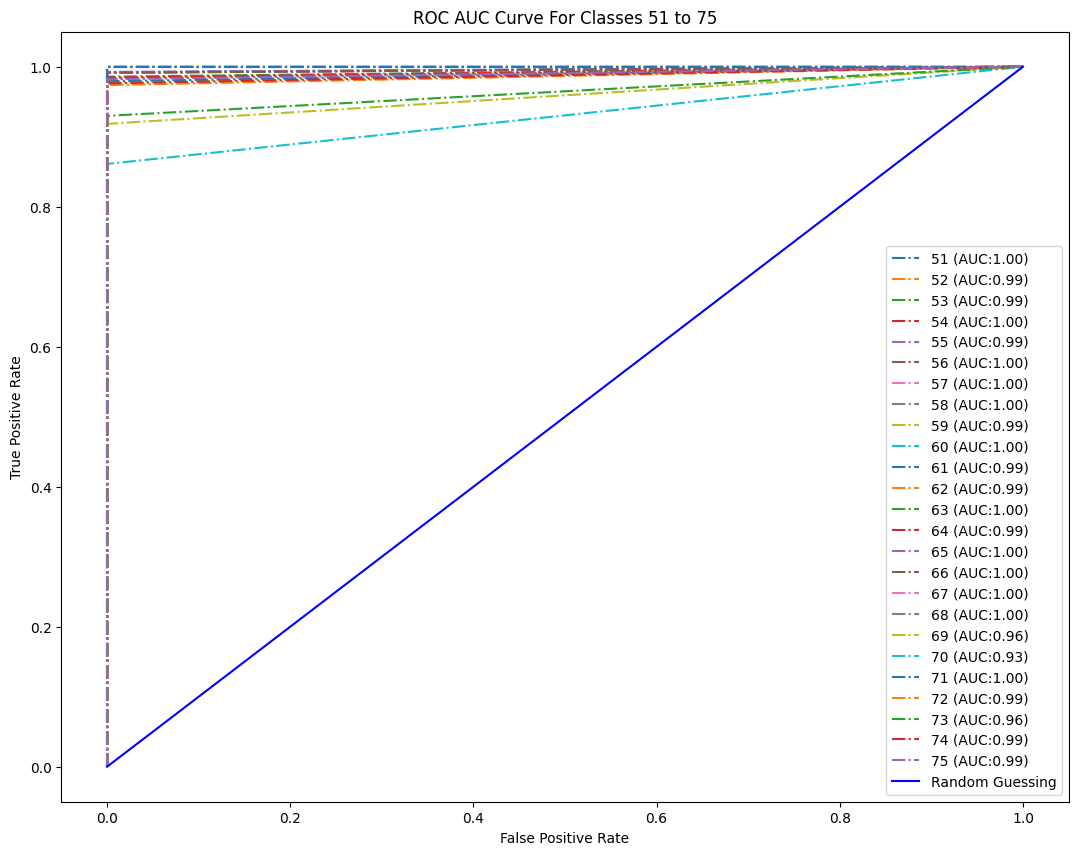}
    \caption{AUC-ROC scores 51-75}
  \end{subfigure}
  \begin{subfigure}[b]{0.48\textwidth}
    \includegraphics[width=\textwidth]{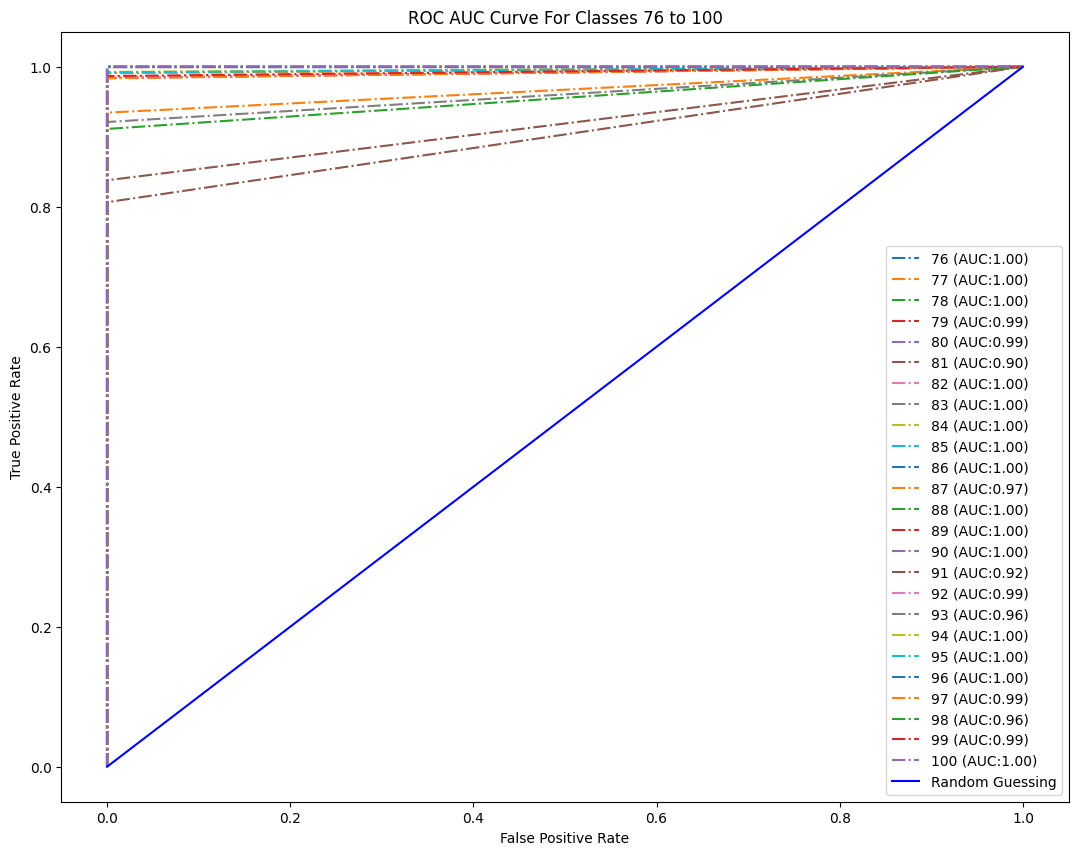}
    \caption{AUC-ROC scores 76-100}
  \end{subfigure}
  \hfill
  \caption{AUC-ROC scores}
  \label{fig:auc_roc_scores}
\end{figure*}

\Cref{fig:our_model_accuracy,fig:our_model_loss} illustrate the accuracy and corresponding loss for both the training and validation datasets. The results indicate a synchronous behavior between validation loss and accuracy, as well as training loss and accuracy, suggesting that ProFamNet avoids over-fitting. The validation loss consistently decreases, and the accuracy increases, with minimal gaps between training and validation metrics. Following model training, predictions on the testing dataset are made, and performance is assessed using F1-score, Precision, and AUC-ROC scores, as defined earlier. Table \ref{tab:top60_protein_family_classes} details the performance metrics for the top $60$ classes, including the number of instances in each class. Additionally, AUC-ROC scores are graphically represented in Figure \ref{fig:auc_roc_scores} for the top $100$ frequent classes, demonstrating that the majority of classification AUC-ROC scores exceed $0.992$, affirming the reliability and stability of ProFamNet. 
Notably, ProFamNet classifier demonstrates commendable performance overall, with no instances of poor performance in any specific class.

\begin{table*}[!]
\centering
\scriptsize  
\begin{tabular}{|l|c|c|c|c|}
\hline
\textbf{Family Class} & \textbf{F1 Score} & \textbf{Precision} & \textbf{Recall} & \textbf{Instances}\\
\hline
\hline
Class-II aminoacyl-tRNA synthetase family	&	0.9929	&	0.9893	&	0.9964	&	559	\\
Class-I aminoacyl-tRNA synthetase	&	0.9780	&	0.9732	&	0.9828	&	407	\\
ATPase alpha/beta chains family	&	0.9887	&	0.9887	&	0.9887	&	354	\\
G-protein coupled receptor 1 family	&	0.9725	&	0.9792	&	0.9659	&	293	\\
Cytochrome b family	&	0.9908	&	0.9889	&	0.9926	&	270	\\
Cytochrome P450 family	&	0.9732	&	0.9621	&	0.9845	&	258	\\
Heat shock protein 70 family	&	0.9745	&	0.9598	&	0.9896	&	193	\\
HisA/HisF family	&	1.0000	&	1.0000	&	1.0000	&	189	\\
MurCDEF family	&	0.9914	&	0.9942	&	0.9885	&	174	\\
Universal ribosomal protein uS4 family	&	0.9943	&	0.9886	&	1.0000	&	173	\\
Chaperonin (HSP60) family	&	0.9848	&	1.0000	&	0.9701	&	167	\\
Universal ribosomal protein uS2 family	&	0.9909	&	0.9939	&	0.9879	&	165	\\
Universal ribosomal protein uS7 family	&	0.9939	&	1.0000	&	0.9878	&	164	\\
Globin family	&	0.9602	&	0.9515	&	0.9691	&	162	\\
Universal ribosomal protein uL16 family	&	0.9873	&	1.0000	&	0.9750	&	160	\\
Universal ribosomal protein uL2 family	&	0.9969	&	0.9938	&	1.0000	&	159	\\
Universal ribosomal protein uS19 family	&	0.9968	&	1.0000	&	0.9937	&	158	\\
Universal ribosomal protein uL14 family	&	0.9936	&	0.9936	&	0.9936	&	156	\\
Universal ribosomal protein uS12 family	&	0.9967	&	1.0000	&	0.9935	&	154	\\
Universal ribosomal protein uS8 family	&	0.9967	&	1.0000	&	0.9934	&	152	\\
Classic translation factor GTPase family, EF-Tu/EF-1A subfamily	&	0.9934	&	1.0000	&	0.9868	&	152	\\
Short-chain dehydrogenases/reductases (SDR) family	&	0.9467	&	0.9530	&	0.9404	&	151	\\
Universal ribosomal protein uL22 family	&	0.9799	&	0.9865	&	0.9733	&	150	\\
Universal ribosomal protein uS11 family	&	0.9899	&	0.9932	&	0.9866	&	149	\\
Universal ribosomal protein uS3 family	&	0.9865	&	0.9932	&	0.9799	&	149	\\
Universal ribosomal protein uS15 family	&	0.9829	&	0.9931	&	0.9730	&	148	\\
Ser/Thr protein kinase family	&	0.8475	&	0.8446	&	0.8503	&	147	\\
Classic translation factor GTPase family	&	0.9896	&	1.0000	&	0.9795	&	146	\\
RNA polymerase beta chain family	&	0.9758	&	0.9860	&	0.9658	&	146	\\
Bacterial ribosomal protein bL33 family	&	0.9896	&	0.9931	&	0.9862	&	145	\\
Krueppel C2H2-type zinc-finger protein family	&	0.9622	&	0.9589	&	0.9655	&	145	\\
Prokaryotic/mitochondrial release factor family	&	0.9965	&	1.0000	&	0.9931	&	144	\\
Bacterial ribosomal protein bL20 family	&	0.9893	&	0.9929	&	0.9858	&	141	\\
Enolase family	&	0.9927	&	0.9927	&	0.9927	&	137	\\
ATPase B chain family	&	0.9963	&	0.9927	&	1.0000	&	136	\\
Universal ribosomal protein uL1 family	&	0.9853	&	0.9781	&	0.9926	&	135	\\
Bacterial ribosomal protein bS18 family	&	0.9888	&	0.9925	&	0.9852	&	135	\\
Universal ribosomal protein uL24 family	&	0.9963	&	0.9926	&	1.0000	&	135	\\
Universal ribosomal protein uL11 family	&	0.9846	&	1.0000	&	0.9697	&	132	\\
Universal ribosomal protein uS10 family	&	0.9728	&	1.0000	&	0.9470	&	132	\\
Universal ribosomal protein uL18 family	&	0.9699	&	0.9556	&	0.9847	&	131	\\
Universal ribosomal protein uL5 family	&	0.9848	&	0.9774	&	0.9924	&	131	\\
Universal ribosomal protein uL3 family	&	1.0000	&	1.0000	&	1.0000	&	130	\\
Bacterial ribosomal protein bS16 family	&	0.9882	&	0.9921	&	0.9844	&	128	\\
Universal ribosomal protein uS13 family	&	0.9845	&	0.9769	&	0.9922	&	128	\\
Complex I subunit 4L family	&	0.9922	&	0.9922	&	0.9922	&	128	\\
Methyltransferase superfamily, RsmH family	&	0.9805	&	0.9692	&	0.9921	&	127	\\
Universal ribosomal protein uL4 family	&	0.9960	&	1.0000	&	0.9921	&	127	\\
SHMT family	&	1.0000	&	1.0000	&	1.0000	&	126	\\
Universal ribosomal protein uL6 family	&	0.9960	&	0.9921	&	1.0000	&	125	\\
Bacterial ribosomal protein bL32 family	&	0.9841	&	0.9688	&	1.0000	&	124	\\
Adenylosuccinate synthetase family	&	0.9960	&	0.9920	&	1.0000	&	124	\\
Universal ribosomal protein uL15 family	&	0.9919	&	0.9919	&	0.9919	&	124	\\
Phosphohexose mutase family	&	0.9801	&	0.9685	&	0.9919	&	124	\\
EF-Ts family	&	0.9879	&	0.9919	&	0.9839	&	124	\\
Elongation factor P family	&	0.9960	&	0.9920	&	1.0000	&	124	\\
SecA family	&	0.9880	&	0.9762	&	1.0000	&	123	\\
\textbf{Average}	&	\textbf{0.9845}	&	\textbf{0.9849}	&	\textbf{0.9843}	&	\textbf{166}	\\
\hline
\end{tabular}
\caption{Top 60 popular protein family classes}
\label{tab:top60_protein_family_classes}
\end{table*}

\subsection{Performance on unreviewed protein sequences}
Motivated by the challenge of identifying uncharacterized proteins, we validated the proposed model on UniProt's unreviewed protein sequences, awaiting full manual annotation. Unlike reviewed records, we intentionally included a broad range of uncharacterized protein sequences, bypassing length-based filtering and utilizing raw data for direct model verification. To address data imbalance, particularly prominent in unreviewed records, we employed micro-F1 metrics for result validation.

The evaluation focused on Human, Rice, A.thaliana, Mouse, and Zebrafish, encompassing diverse species. Across these organisms, F1-scores ranged from 0.7424 to 0.8982, highlighting the model's effectiveness on unreviewed data and its comparability to automatic annotation methods. The results underscore the model's robust performance across varied species and uncharacterized protein sequences, further validating its potential utility. A comparison with the state of the art work of Da Zhang et al. \citep{zhang2020protein} is also given indicating that ProFamNet also outperforms it except for mouse species. The dataset statistics, detailed in Table \ref{tab:unreviewed_performance}, showcase the diversity and volume of the unreviewed records used for validation.

\begin{table}[t]
\centering
\begin{tabular}{||l|c|c|c||}
\hline
\textbf{Species} & \textbf{Instances} & \textbf{F1 Score} & \textbf{F1 Score}\\
 &  & \citep{zhang2020protein} & \textbf{ProFamNet}\\

\hline\hline
Human & 6,489 & 0.8534 & \textbf{0.8708}\\
\hline
Rice & 2,297 & 0.6482 & \textbf{0.7424}\\
\hline
A.thaliana & 1,921 & 0.7537 & \textbf{0.8208}\\
\hline
Mouse & 4,302 & \textbf{0.8393} & 0.8334 \\
\hline
Zebrafish & 2,586 & 0.8573 & \textbf{0.8982}\\
\hline
\end{tabular}
\caption{Performance on unreviewed protein sequences}
\label{tab:unreviewed_performance}
\end{table}

\section{Conclusion}
In this study, we establish a fusion of 1D-CNN, Bi-LSTM, and an attention network for predicting protein families solely from primary protein sequences. In contrast to methods that manually generate and select features, ProFamNet utilizes the raw data of primary protein sequences, eliminating the need for manual feature engineering. The protein sequences in the proposed study are embedded as matrices with predefined lengths. Using kernels of various sizes on each matrix, we automatically extract diverse temporal features. Post feature extraction, fully connected layers are employed for the classification of protein families. Experimental results demonstrate that ProFamNet surpasses most existing approaches without relying on human-designed features or generating N-Gram dictionaries.

As this research is motivated by the classification of uncharacterized proteins, we apply ProFamNet to the most popular unreviewed protein sequences in the UniProt database—proteins that are automatically annotated without undergoing review. The achieved F1-score of 0.8708 attests to the predictive capability of the proposed model. An inherent extension of this work involves predicting protein interactions based on these embeddings, aiming to demonstrate whether the embeddings can successfully reconstruct protein interactions. \\

\textbf{Declaration of Competing Interest} 

The authors declare no competing interests.
\\

\textbf{Acknowledgement}

This work is partially supported by faculty research
support grant, National University of Computer and Emerging Sciences, Pakistan.



\bibliographystyle{elsarticle-harv}
\bibliography{PBLSTM_References}


%
%
%
\end{document}